\documentclass{sig-alternate}
\usepackage{multirow}
\usepackage{url}

\begin{document}
%
\conferenceinfo{ACM Workshop on Recommendation Systems for Television and Online Video}{'2014 Foster City, California USA}

\title{A Mood-based Genre Classification of Television Content
\titlenote{This work was supported by Science Foundation Ireland through the CLARITY Centre for Sensor Web Technologies under grant number 07/CE/I1147 and through the Insight Centre for Data Analytics under grant  SFI/12/RC/2289.}}

\numberofauthors{2} 
%
\author{
%
%
\alignauthor
Humberto Jesús Corona Pampín\\
       \affaddr{Insight Centre for Data Analytics}\\
       \affaddr{University College Dublin, Ireland}\\
             \email{humberto.corona@insight-centre.org}
\alignauthor
Michael P. O'Mahony\\
          \affaddr{Insight Centre for Data Analytics}\\
       \affaddr{University College Dublin, Ireland}\\
             \email{michael.omahony@insight-centre.org}
}

\maketitle
\begin{abstract}

The classification of television content helps users organise and navigate through the large list of channels and programs now available. In this paper, we address the problem of television content classification by exploiting text information extracted from program transcriptions. We present an analysis  which adapts a model for sentiment that has been widely and successfully applied in other fields such as music or blog posts. We use a real-world dataset obtained from the Boxfish API to compare the performance of classifiers trained on a number of different feature sets. Our experiments show that, over a large collection of television content, program genres can be represented in a three-dimensional space of valence, arousal and dominance, and that promising classification results can be achieved using features based on this representation. This finding supports the use of the proposed representation of television content as a feature space for similarity computation and recommendation generation. 

\end{abstract}




\section{Introduction}

The problem of choice overload in television is well known. For example, in 2009 there were 2,218 television broadcast stations in the U.S.\footnote{\url{http://en.wikipedia.org/wiki/List_of_countries_by_number_of_television_broadcast_stations}, Accessed 20/07/2014.}, making it very difficult for users to manually organise, browse or decide what content is relevant or best suited for them. Thus, there is a need for new ways to classify television content that can be applied by intelligent systems to enable content discovery. 

Most of the research on television content classification is based on audio and visual features, focusing on genre classification \cite{Eggink2012, Jasinschi2001} and on the relationship between content and industry, audience, and culture \cite{Mittell2001}. However, in this paper we focus on the analysis of text extracted from television program scripts for genre classification.  We use metadata obtained from the Boxfish API\footnote{\url{http://boxfish.com/api}} to  build a textual representation of television program and channel content. This allows us to explore content in a three-dimensional space of affect defined by valence, arousal and dominance. To this end, we follow the approach presented in \cite{Dodds2009} and expand it by also considering the arousal and dominance dimensions and by applying this representation in the context of genre classification in the television domain.

The paper is organised as follows. First, Section \ref{section:related} describes related work. Then, Section \ref{section:datasets} introduces the datasets used in our study. Section \ref{section:exp1} describes a feature analysis over a year of television content, and presents a mood analysis for one particular news channel. Section \ref{section:exp2} introduces our classification approach, and discusses the results obtained. Finally, Section \ref{section:conc} presents conclusions and future work.


\section{Related Work}\label{section:related}

The moods associated with multimedia content such as television programs or songs are difficult to infer: people perceive them differently \cite{Song} and they are culture dependent \cite{Kosta2013}. For example, songs like \emph{Bohemian Rhapsody} by \emph{Queen} are non-trivial to classify in this dimension. 

Several ontologies for mood classification rely on models developed in the psychology field, \emph{Rusell's model of affect} \cite{Russell1980} being one of the most widely used. This model represents each mood in a two-dimensional space defined by valence $v$ (which measures the good--bad dimension of sentiment) and arousal $a$ (which measures the active--passive dimension of sentiment). Thus, each mood $m \in M = \left\{v,a\right\}$ can be represented by a vector in this two-dimensional space. The model is based on the evidence that the affective dimensions are built in a \emph{highly systematic fashion}, instead of being independent dimensions.

Dodds et al.  \cite{Dodds2009} uses features extracted from lyrics and the Affective Norms for English Words (ANEW) dataset \cite{Bradley1999} to measure the average happiness (valence) of songs, blogs and State of the Union presidential speeches. The aim of the work is to quantify the evolution of the overall happiness in different contexts. The approach  calculates the average valence of each document (song, blog post or speech) by counting the number of times each of the terms in the ANEW dataset appears in the document, and multiplying it by its associated mean valence value. The results show that, for example, valence can help distinguish between music genres when a large number of songs are considered, and that interesting trends in presidential speeches are revealed.

Eggnik et al. \cite{Eggink2012} perform a large scale experiment on the mood classification of television programs from the BBC channel using a live user study. Participants were asked to watch short clips from television programs and assign mood labels. The results obtained showed that there was consensus on the mood labels applied, and an automatic classification based on the data obtained a 90\% accuracy for certain programs. Moreover, the study performed a principal component analysis, finding two main components in the mood of television content: one related to the seriousness of the program and the second related to the perceived pace.

A mood-based similarity metric to exploit movie mood similarities for context-aware recommendations is presented in \cite{Shi2010}. Here, the proposed metric is used in  a joint matrix factorisation model, obtaining results that lead to better recommendations compared to other mood-based movie similarities considered (in the context of mood-based recommendations).

From the related work it is clear that the analysis of television content using a multidimensional mood space is an interesting problem. Thus, in this paper we consider a text-based classification of television content using features based on the dimensions that define Rusell's model, and also consider the dominance (or control) dimension. Moreover, we follow the approach proposed in \cite{Dodds2009} expanding the study to the valence and dominance dimensions, (in line with \cite{Hu2010xx}).

\section{Datasets}\label{section:datasets}

The \emph{ANEW} dataset \cite{Bradley1999} is a collection of  2,476 words annotated with emotional ratings in three dimensions --- \emph{valence}, \emph{arousal} and \emph{dominance}. The dataset was created using human assessment, and it aims to provide a \emph{set of normative emotional ratings} for the words included. For each dimension, the dataset contains the mean and standard deviation of the ratings values obtained for each word. Here, we normalise the original $[1-9]$ scale to $[0-1]$. 

All  the television content information was obtained through the Boxfish API, which provides the electronic programming guide for various channels and the genre associated with each program. We describe the data obtained in detail below.

\begin{itemize}

\item We selected eight different channels which capture a broad range of programs and genres; news (\emph{FOX News, CNN, MSNBC}), general entertainment (\emph{FOX, E!}), Science Fiction (\emph{SyFy}), educational (\emph{Discovery Channel}) and children (\emph{Cartoon Network}). For each channel, we obtained the electronic programming guide, which contains the program schedule for each of the selected channels, including program title, showtime and  program genre.

\item Using the Boxfish API, we selected those genres for which at least 20 programs were available. These were \emph{reality, documentary, animated, newscast} and \emph{horror}.  

\item The terms derived from program transcriptions were also obtained using the API. For each item (program or channel) considered, we queried the total number of occurrences of each term contained in the ANEW dataset over a period of time. On average, there were 283 and 1,569 distinct terms per program and per channel, respectively. Overall, 2,034 distinct terms were obtained for all programs and channels considered, out of the 2,476 included in the ANEW dataset.

\end{itemize}

\section{Feature Analysis}\label{section:exp1}

In this section, we analyse the content of different television channels and their relationship with the valence, arousal and dominance dimensions. We use the approach proposed in \cite{Dodds2009} to calculate the average valence, and we expand it by also considering the arousal and dominance dimensions. Thus, a television program is defined by its mood $m \in M = \left\{v,a,d\right\}$  in this three-dimensional valence-arousal-dominance space.  With the proposed approach we expect to understand to what extent these dimensions can distinguish between different kinds of television content. 

\subsection{Methodology} 

We perform the analysis over a year of television content, from January to December 2013. We obtain the count of ANEW terms for each week for each of the selected channels using the \emph{keyword mentions endpoint} of the Boxfish API. The mean valence (arousal and dominance) values are calculated by multiplying the number of times a term occurs by its associated valence (arousal and dominance) value in the ANEW dataset, as in previous work \cite{Dodds2009}.

\subsection{Results} \label{sec:feature_results}
\begin{figure}[h]   
  \centerline{
 \includegraphics[width=0.9\columnwidth]{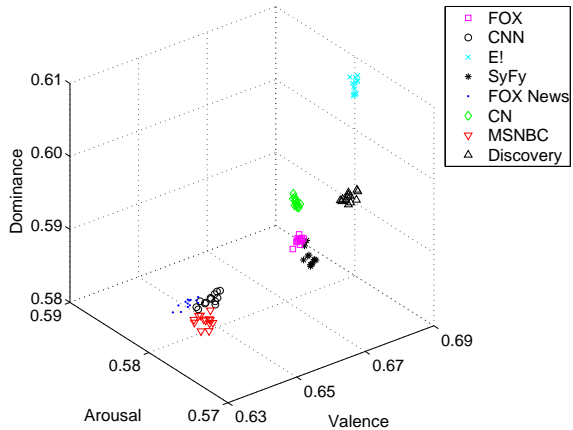}}
  \caption{Valence, arousal and dominance for a year of television broadcasting per channel.}
   \label{fig:vaGenres}
\end{figure}

Figure \ref{fig:vaGenres} presents, for each channel, the mean valence, arousal and dominance values calculated over thirteen periods of four weeks' duration\footnote{Taking averages over periods of four weeks is performed for clarity of presentation. Moreover, these periods correspond to approximately one month of television content.}. From the results we can infer that the valence and dominance dimensions, in particular, can potentially help to classify television content by channel (at least for the channels considered here). For example, compared to other channels, \emph{E! Entretaniment} has on average a much higher valence (0.684) --- it can be considered a very happy channel --- and dominance (0.608), which correlates with the nature of this channel's content (mainly focused on general entertainment and reality television). Moreover, all the news channels are clustered together in this space, showing the lowest mean valence and dominance values (for example, 0.639 and 0.585 for \emph{FOX News}, respectively). These values also appear to be well correlated to the typical content of news channels and the language used. Finally, it can be seen that the rest of the entertainment channels are also clustered together in the space. 

It is also important to analyse the deviation over these mean values, as shown in Table \ref{table:stats}. For example, in the \emph{Cartoon Network}, (CN) channel, the mean standard deviation of valence over the 52 weeks is 0.179. This relatively high variation is due to the fact that the channel broadcasts a wide range of shows, intended for different audiences groups.  For example, a particular episode of \emph{The Amazing World of Gum Ball} has a mean valence of 0.610, which is lower than average for this channel. However, this is understandable as this show contains some references to sex and nudity, violence and profanity\footnote{\url{http://www.imdb.com/title/tt1942683/parentalguide}}. On the other hand, an episode of \emph{Grobjad}, which contains no references to mature content\footnote{\url{http://www.imdb.com/title/tt2406986/parentalguide}}, has a mean valence of 0.690. Moreover, there is no intersection between the top-6 most similar programs to each of the two examples considered, based on data obtained from IMDB. Thus, while the mean valence, arousal and dominance values computed over all programs broadcast by a channel appear to be discriminative, classifying individual programs by channel (i.e. by genre) may be problematic.

\begin{table}[h!]
\centering
\resizebox{\columnwidth}{!}{%
\begin{tabular}{|c|c|c|c|}
\hline
   Channel 	& Valence & Arousal & Dominance  \\ \hline \hline
E!  			&0.684    (0.168)    &0.577    (0.098)    & 0.608    (0.100)  \\ 
Discovery 		&0.665    (0.169)    &0.570    (0.098)    & 0.601    (0.103)  \\ \hline
CN   			&0.663    (0.179)    &0.575    (0.101)    & 0.597    (0.105)  \\ \hline
SyFy  		&0.654    (0.175)    &0.570    (0.099)    & 0.595   (0.108) \\ 
FOX        		&0.658    (0.171)   & 0.573    (0.009)  &  0.594    (0.105)       \\ \hline
FOX News   	&0.639    (0.177)    &0.579    (0.097)    & 0.585    (0.109)                \\ 
CNN  		&0.643    (0.176)   & 0.577    (0.098)   & 0.586    (0.108)   \\ 
MSNBC  		&0.635    (0.177)    &0.575    (0.098)    & 0.586    (0.110)     \\ \hline
\end{tabular}
}
\caption{Valence, arousal and dominance mean and standard deviation (in parentheses) values per television channel.}
\label{table:stats}
\end{table}

\begin{figure}[h]   
  \centerline{
 \includegraphics[width=0.9\columnwidth]{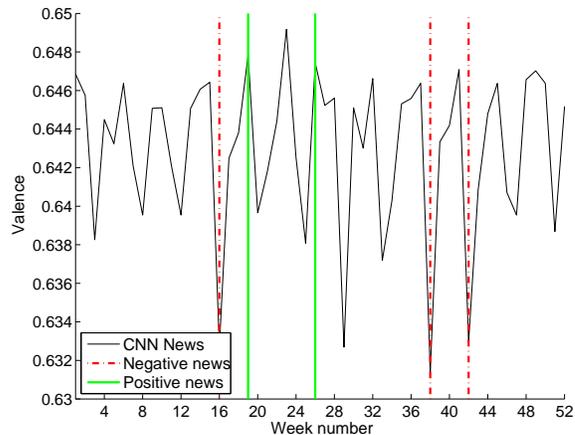}}
  \caption{Valence per week  of \emph{CNN} content in 2013.}
   \label{fig:vNews}
\end{figure}

In Figure \ref{fig:vNews} we present a valence analysis of CNN news channel content by week over one year (2013). The vertical markers highlight some of the top news events\footnote{\url{http://www.infoplease.com}} from the year. For example, low valence values are seen for events such as the Boston Marathon bombings (week 16), the Navy Yard Shooting (week 38) or the collapse of Obamacare (week 42). We also found high  valence values correlated with news events such as the liberation of three women kidnaped in Ohio (week 19), and the Supreme Court ruling on the DOMA and California Marriage Equality Bill (week 26).

The results presented in this analysis show that the valence and dominance dimensions, in particular, can be used to distinguish between the content broadcast by different channels. However, as mentioned previously, classifying the genre of individual television programs (for use in a recommender system) based on these dimensions may present challenges (given the variance observed within genres), an analysis of which is considered in the next section.


\section{Program Genre Classification}\label{section:exp2}
In this section, we analyse the performance of a single-label supervised classification approach for television genre classification using the selected genre categories derived from the Boxfish API (Section \ref{section:datasets}). We use an instance-based representation of each television program based on  statistical features derived from the valence, arousal and dominance dimensions described above. In particular, we consider an early-fusion ensemble approach \cite{IanH.Witten2005} in which all these meta-features are combined into a single feature space. We compare this approach against a standard \emph{vector space model} (VSM) (based on terms in program transcriptions which are also present in the ANEW dataset) approach.

\subsection{Classification Approach}
Feature-based instances for each television program are created as follows. First, we select all the television programs from each of the genres and channels described in Section \ref{section:datasets}.  For each program, we calculate mean valence  (likewise mean arousal and dominance) values over all terms contained in both the program transcriptions and the ANEW dataset. Table \ref{table:metafeatures} presents the features used in this study. 

\subsection{Experimental Methodology}
In total we obtained 343  television programs, queried over a period of two weeks in February 2014. The distribution of genres in program instances was as follows: \emph{animated (120)}, \emph{documentary (65)}, \emph{horror (24)}, \emph{newscast (41)} and \emph{reality (93)}. The classification was performed using the Weka machine learning framework \cite{IanH.Witten2005}. A standard 5-fold cross validation approach was used to evaluate performance, expressed in terms of true positive (TP) rate, false positive (FP) rate and area under the ROC (AUC) for each class. Both the VSM and meta-features approaches were evaluated using a Na\"{\i}ve Bayes classifier. 

\begin{table}[h!]
\resizebox{\columnwidth}{!}{%
\begin{tabular}{|c|l|}
\hline
\textbf{Feature Group}                                                                                 & \multicolumn{1}{|c|}{\textbf{Feature}} \\ \hline \hline
\multirow{5}{*}{\begin{tabular}[c]{@{}c@{}}ANEW features\\ (valence, arousal, dominance)\end{tabular}}  
                                                                                                       & Minimum value                       \\ \cline{2-2} 
                                                                                                       & Maximum value                         \\ \cline{2-2}
                                                                                                       & Mean value \\ \cline{2-2} 
                                                                                                       & Standard deviation                    \\ \cline{2-2} 
                                                                                                       & Median value                           \\ \hline
\multirow{4}{*}{Stylistic features}                                                                    & Num. words                        \\ \cline{2-2} 
                                                                                                       &  Num. unique words                 \\ \cline{2-2} 
                                                                                                       &  Num. unique ANEW words            \\ \cline{2-2} 
                                                                                                       & Max. word frequency                     \\ \hline

\end{tabular}}
\caption{Meta-feature representation.}
 \label{table:metafeatures}
\end{table}
\vspace{-2mm}

\subsection{Results}

The results provided by each classifier are shown in Table~\ref{table:classification}. In general, VSM provided better accuracy compared to the meta-features approach. The exception to this trend was the \emph{horror} genre, where the meta-features classifier obtained an AUC score of 0.865 compared to 0.649 for VSM. For both approaches, the best classification accuracy was achieved in respect of the \emph{newscast} genre, with very high AUC scores $(> 0.9)$ observed in both cases. This finding indicates that a relatively distinct vocabulary is associated with this genre in particular. Moreover, this result is inline with that of Section \ref{sec:feature_results}, where news channel content was characterised by especially low valence and dominance values (Figure \ref{fig:vaGenres}).

\begin{table}[h]
\resizebox{\columnwidth}{!}{%
\begin{tabular}{|l|c|c|c|c|c|c|}
\hline
\multicolumn{1}{|c|}{\multirow{2}{*}{\textbf{Genre}}} & \multicolumn{3}{c|}{\textbf{VSM}}        & \multicolumn{3}{c|}{\textbf{Meta-features}} \\ \cline{2-7} 
\multicolumn{1}{|c|}{}                                & \textbf{TP} & \textbf{FP} & \textbf{AUC} & \textbf{TP}  & \textbf{FP}  & \textbf{AUC} \\  \hline \hline
Animated                                              & 0.725       & 0.004       & 0.942        & 0.792        & 0.161        & 0.885        \\ \hline
Documentary                                           & 0.554       & 0.025       & 0.870        & 0.447        & 0.144        & 0.780        \\ \hline
Horror                                                & 0.208       & 0.013       & 0.649        & 0.583        & 0.107        & 0.865        \\ \hline
Newscast                                              & 0.976       & 0.053       & 0.972        & 0.707        & 0.056        & 0.905        \\ \hline
Reality                                               & 0.882       & 0.260       & 0.845        & 0.333        & 0.064        & 0.729        \\ \hline \hline
Weighted Average                                               & 0.729       & 0.084       & 0.885        & 0.583        & 0.115        & 0.824        \\ \hline
\end{tabular}
}
\caption{Genre classification performance.}
\label{table:classification}
\end{table}

\vspace{-2mm}

Although outperformed by VSM in most cases, the meta-features based approach shows promising results and offers a new representation for this type of data. In particular, the meta-features classification results are different in terms of their relative ordering (by genre) according to accuracy, indicating that the two approaches work on different aspects of the data. Thus, an ensemble technique, combining both types of features, may provide enhanced performance. An analysis of such a technique is left to future work.

\section{Conclusions and Future Work}\label{section:conc}

In this paper we  have presented a study of television content classification, relying solely on textual features as a source of information. From the feature analysis described in Section \ref{section:exp1}, it is clear that television content, at least at a high (i.e. channel) level, can be discriminated by the proposed three-dimensional space of affect. While classifying the genres of individual television programs using this approach in general did not outperform a traditional VSM based classifier, nevertheless there is evidence to suggest that meta-features based on valence, arousal and dominance values have the potential to contribute to enhanced classification performance, particularly if used in combination with other feature types. Moreover, we note that the meta-features used in this work were based on ANEW values computed over program transcription terms (referred to as ``keywords'') returned by the Boxfish API; although 82\% of these terms were present in the ANEW dataset, better performance may be achieved if  complete program transcription texts were available. In future work, we will also consider using the model of affect in a personalised content-based recommendation approach, as well as conducting live user studies to understand how individuals respond to mood-based recommendations\footnote{Additional information (including source code) for this paper can be found at \url{https://github.com/hcorona/recsystv-2014}}.


\balancecolumns

\end{document}